\documentclass[letterpaper]{article} 
\usepackage{aaai24}  
\usepackage{times}  
\usepackage{helvet}  
\usepackage{courier}  
\usepackage[hyphens]{url}  
\usepackage{graphicx} 
\usepackage{natbib}  
\usepackage{caption} 
\usepackage{algorithm}
\usepackage{algorithmic}

\usepackage{xcolor}
\usepackage{fontawesome}
\usepackage{soul}

\usepackage{xspace}

\usepackage{tcolorbox}
\usepackage{tabularx}

\usepackage{lipsum}
\usepackage{multirow}
\usepackage{amsfonts}
\usepackage{amsmath}
\usepackage{tikz}

\usepackage{newfloat}
\usepackage{listings}
\DeclareCaptionStyle{ruled}{labelfont=normalfont,labelsep=colon,strut=off} 
\floatstyle{ruled}
\newfloat{listing}{tb}{lst}{}
\floatname{listing}{Listing}
\title{A Conceptual Framework for Ethical Evaluation of Machine Learning Systems}

\author{
    Neha R. Gupta\textsuperscript{\rm 1}, 
    Jessica Hullman\textsuperscript{\rm 2}, 
    Hari Subramonyam\textsuperscript{\rm 3}
}
\affiliations{
    \textsuperscript{\rm 1} Carnegie Mellon University, Pittsburgh, USA\\
    nehagupt@andrew.cmu.edu\\
    \textsuperscript{\rm 2} Northwestern University, Evanston, USA\\
    jhullman@northwestern.edu\\
    \textsuperscript{\rm 3} Stanford University, Stanford, USA\\
    harihars@stanford.edu
}

\begin{document}

\maketitle

\begin{abstract}
Research in Responsible AI has developed a range of principles and practices to ensure that machine learning systems are \emph{used} in a manner that is ethical and aligned with human values. However, a critical yet often neglected aspect of ethical ML is the ethical implications that appear when \emph{designing evaluations} of ML systems. For instance, teams may have to balance a trade-off between highly informative tests to ensure downstream product safety, with potential fairness harms inherent to the implemented testing procedures. We conceptualize ethics-related concerns in standard ML evaluation techniques. Specifically, we present a utility framework, characterizing the key trade-off in ethical evaluation as balancing information gain against potential ethical harms. The framework is then a tool for characterizing challenges teams face, and systematically disentangling competing considerations that teams seek to balance. Differentiating between different types of issues encountered in evaluation allows us to highlight best practices from analogous domains, such as clinical trials and automotive crash testing, which navigate these issues in ways that can offer inspiration to improve evaluation processes in ML. 
Our analysis underscores the critical need for development teams to deliberately assess and manage ethical complexities that arise during the evaluation of ML systems, and for the industry to move towards designing institutional policies to support ethical evaluations.
\end{abstract}

\section{Introduction}
Machine learning (ML) model evaluation typically focuses on estimating errors of prediction or estimation via quantifiable metrics. Given the increasing size and complexity of ML systems, comprehensive evaluations should ideally be multifaceted. For example, evaluations of large ML systems may include several methods, including A/B testing on live populations, adversarial testing to produce undesirable outputs, and comprehensive audits documenting outputs. 

Potential ethical harms of ML systems have gained increasing attention in the broad Responsible AI community. However, even when evaluation metrics are expanded beyond performance to include factors like fairness, privacy loss, or other harms induced by the machine learning system, this is often focused on the ethical harms of the released system, overlooking possible harms incurred during the machine learning development lifecycle itself. This is problematic because evaluation approaches do have the potential to cause ethical harm during evaluation. In a noteworthy example, Tesla's autonomous vehicle live testing systems on real roadways in California, has been widely criticized for being involved in various crashes \cite{nayakbloomberg}. 

\textit{How should practitioners evaluate large complex systems with potentially unknown ethical harms across the engineering lifecycle, including during the evaluation process?} We provide a conceptual framework that casts the primary trade-off in ethical evaluation decision-making as balancing the goal of optimizing for information gained in an evaluation, against the possible ethical harms that are induced. 

Based on our sketch of this fundamental problem that practitioners face, we identify a series of challenges that can cause practitioners to stumble in selecting ethical evaluation practices.
We illustrate these challenges using real-world examples of machine learning evaluations that encountered them. Then, we draw parallels between these challenges and evaluation practices in domains other than machine learning, to explore potential mitigation techniques. Together, our conceptual framework and characterization of challenges are intended to stimulate discussion among researchers and evaluation teams on how to balance information gain with potential ethical harm, and to motivate future exploration of policies or best practices for machine learning evaluation.
\section{Related Works}
\subsection{Ethical AI}

A growing body of literature discusses properties that ethical machine learning systems should inherently possess, and provides principles and guidelines for testing~\cite{jobin2019global,zhang2020machine, martinez2022software}.  Broadly speaking, the ethical values identified by prior work include: (1) \textbf{Non-maleficense}, which measures the extent to which the evaluation workflows and outcomes do not inflict harm or injuries on any individual or population~\cite{mehrabi2021survey}. (2) \textbf{Privacy}, which as a value refers to the principle of protecting personal and sensitive information from unauthorized access, use, or exposure during the entire ML lifecycle~\cite{liu2021machine}. (3) \textbf{Fairness}, in which the goal is to achieve equitable treatment and outcomes for all individuals, ensuring that the benefits and burdens of AI systems are distributed justly across diverse populations~\cite{mehrabi2021survey}, (4) \textbf{Cultural Sensitivity}, which involves designing algorithms and models that are attuned to and respectful of cultural differences, ensuring that they do not perpetuate stereotypes, biases, or insensitivities~\cite{jo2020lessons}. (5) \textbf{Sustainability}, which ensures that models are developed in a manner that is environmentally responsible, economically feasible, and socially equitable~\cite{eugoals,lacoste2019quantifying}. (6) \textbf{Societal Impact}, which refers to ensuring that the models contribute positively to societal well-being, address social issues, and do not harm individuals or communities~\cite{birhane2022values}. We emphasize that this is not a comprehensive listing of all ethical values that a machine learning system may seek to accomplish. We provide examples of these categories when characterizing challenges in designing ethical evaluations in later sections.

\subsection{ML System Evaluation Practices}
An \emph{evaluation} is the process by which practitioners detect differences between desired and actual model behavior \cite{zhang2020machine}, through empirical assessment of model properties \cite{shevlane2023model}. A growing body of work creates more comprehensive methods with which to evaluate systems, rather than providing a singular empirical metric or set of metrics. Some evaluation methods can be conducted \textit{pre-deployment}, such as A/B testing or live testing. Other mechanisms are used \textit{post-deployment}, such as bug bounty challenges, and provide infrastructure to support stakeholder feedback. A particular evaluation process may involve choosing one or many evaluation metrics to measure. These decisions are critical because they impact actions that are taken post-evaluation to improve system capabilities. They also may be associated with the potential for ethical harm incurred in the evaluation process, or after product release. \emph{We define an ethical evaluation as an evaluation that does not sacrifice ethical values in its implementation, and attempts to forecast downstream ethical harms across the product lifecycle.} 

Finally, by considering how to conceive of the value of information about model performance gained through an evaluation, our work is related to data valuation.
Prior theoretical work in machine learning and related fields studies the value of data for purposes like explainability (e.g., the Shapley framework~\cite{ghorbani2019data}), data markets and incentivizing collaboration in ML~\cite{castro2023data,sim2020collaborative}, and value of accurate or improved prediction for goals like treatment assignment or welfare maximization~\cite{liu2024actionability,perdomo2023relative}.

\section{Ethical Evaluation Model}
\label{sec:framework}
\subsection{Motivation}
The economics discipline has a long history of creating highly simplified models of complex real-world processes to assist with predicting the consequences of actions. Abstracting away non-essential features of the complex real-world permits systematic reasoning. 

For example, economic policy-makers concerned with pricing wheat might use a simplified model that includes the costs to the farmer while abstracting away other potentially relevant characteristics, such as soil quality and his educational background \cite{friedman1953methodology}. 
Our conceptual model of the key trade-off in ML practitioners' evaluation decision-making focuses on the value of information gain relative to ethical harms. 
However, rather than contributing new theoretical results, our goals are epistemological: to prompt reflection on what it would mean to select the best evaluation in a way that accounts for potential ethical harms induced in evaluation.
By conceptualizing the idea of an optimal balance between competing concerns in designing ML evaluations, our framework is meant to highlight difficult questions that largely remain un-navigated in the literature and practice of ML evaluation design, rather than to imply that a normative evaluation design is easy to identify. 
Below, we discuss the implications of components of the model, including the acceptability of some of the assumptions made for the sake of this model, issues that arise due to differing aims, and the subjectivity of variables in further sections.

\subsection{Model Properties}
\textbf{ML teams select from a space of possible evaluations.}

An evaluation is a protocol for assessing and measuring a model's performance against a set of defined criteria or benchmarks, including specification of which information to collect and how. ML development teams face various considerations when planning evaluations that involve complex decisions across evaluation scope, context, and effect~\cite{zhang2020machine, riccio2020testing, song2022exploring}; prior work has described how practitioners can suffer from a ``paradox of choice'' when it comes to deciding how to perform evaluation~\cite{goel2021robustness}. 
We represent the space of possible evaluations under consideration by a team as $A = \{a_1,a_2,\ldots \}$ where $a$ is an evaluation decision (e.g., evaluation method, metrics, sample selection, etc.). 
$$\textit{choose some a}\in A$$

\vspace{3mm}
\noindent\textbf{The utility of an evaluation approach depends on the relative value of information gained, ethical harms, and resource costs.}
The fair ML literature has represented decisions about model choice in ML in a utility framework, where models provide utility as a function of costs and benefits  \cite{corbett2017algorithmic, corbett2018measure, chohlas2021learning}. For example, in \citeauthor{corbett2017algorithmic} \shortcite{corbett2017algorithmic}, the authors conceptualize `immediate utility' reflecting the costs and benefits of a fair decision by a policymaker in the setting of pre-trial bail release decisions. A utility framing is also used in \citeauthor{hutchinson2022evaluation}, to illustrate the task of evaluating an ML model's suitability for use in a specific application ecosystem.

Our conceptualization similarly draws on a utility framework common in statistical decision theory~\cite{savage1972foundations,steele2015decision,von2007theory}, but expands this to a broader view applied to decisions made by teams evaluating ML models or systems.
The proposed utility function implies that each evaluation decision option the team is considering can be compared along a single dimension (utility) along which they can be ranked.

We conceive of three categories of inputs to the utility function. The first concerns the value of information gain. The information learned from a test, regarding differences between desired and actual model behavior~\cite{zhang2020machine}, implies a gain from the evaluation process. Information gained in conventional ML performance evaluation includes estimates of how well a trained model generalizes to new samples from the same distribution the model was trained on, as well as measures of the robustness of the model, i.e., the degree to which the model or system maintains its correctness and performance under varying conditions and inputs, including invalid or adversarial inputs~\cite{tjeng2017evaluating,zhang2020machine}. 
When the evaluation produces information about the model performance along ethical dimensions, the information gained may come in the form of the expected magnitude or frequency of harms upon deployment. 

The second component is ethical harms of the evaluation. Often overlooked, ethical issues incurred during evaluations, or downstream ethical issues not adequately predicted (and consequently encountered after deployment), can diminish the acceptability of the results, and the overall value and utility of the information derived from the evaluation.

The third component measures the material resources required to conduct an evaluation. Teams often consider options for reducing the costs of tests via methods such as test prioritization, in which inputs generated for tests are limited to inputs that are most indicative of problematic behavior~\cite{zhang2020machine}. Costs can take several forms. For instance, monetary constraints may restrict data collection abilities including the number of labelled data annotations procured for supervised ML models \cite{liao2021towards, goel2019crowdsourcing}. Cost constraints through labor force availability and time constraints can shift teams towards using automated software testing \cite{dustin2009implementing}.
These resource constraints can challenge responsible model development~\cite{hopkins2021machine}.

Consequently, we represent the utility of an evaluation decision 
as having these three inputs, with the information gain representing gains to utility and potential ethical harm and material cost representing decreased utility:

\begin{multline*}
       u(a) = (\textit{information gain} - 
       \textit{ethical harm}- \textit{cost}) 
\end{multline*} 

\vspace{3mm}
\noindent\textbf{Information gain, ethical harm, and cost are forecasts}. 

Before conducting an evaluation, a team cannot precisely predict the information gained about model performance, potential ethical harm, or exact material costs involved. Consequently, information gain, ethical harm, and cost as predicted values, which we represent as expectations over relevant sources of randomness. The fact that these quantities must be predicted emphasizes the uncertainty under which evaluation decisions are necessarily made.

These values are represented as expectations over distribution of possible values. Estimating these distributions is a fundamental part of the challenge in selecting an evaluation. Any particular evaluation involves a sample of instances for which evaluation data is gathered. Many evaluations sample from a population of participants. Complications arise as models can differ in social impact across groups, notably underrepresented groups \cite{hutchinson2022evaluation}. Thus, the estimate of expected ``ethical harm" requires averaging harm across several individuals who experience disparate impact from the models.
Moving forward, we abbreviate information gain as $IG(a)$, and ethical harms as $EH(a)$ :

$$u(a) = \mathbb{E}(\textit{IG}(a)) - \mathbb{E}(\textit{EH}(a)) - \mathbb{E}(\textit{cost}) $$
\noindent\textbf{Ethical harm can be decomposed by distinct ethical values.}
It is important to differentiate between various ethical values in our model, because it has been established that there are situations where some models may benefit a particular ethical value at the cost of another. For example, a privacy and fairness trade-off affects some ML models \cite{pujol2021equity}. While interactions between ethical concerns may exist, for simplicity we think of $\mathbb{E}(EH(a))$ as representing a weighted sum of various ethical values, so $\mathbb{E}(EH(a))=\sum_{j}w_j\mathbb{E}(EH_j(a))$ where $j$ paramaterizes the ethical values discussed in Section 2. These weights can represent differing ethical priorities of teams or regulatory requirements on particular values.

$$ \mathbb{E}(EH(a)) = \sum_{j}w_j \mathbb{E}(\textit{EH}_j(a)) $$

\noindent\textbf{The best evaluation method has the highest utility.}
We represent the optimal choice of evaluation practice for the team as the decision with the highest utility. An evaluation approach $a$ equals the optimal decision $a^*$ if it is the utility-maximizing decision, indicated as:

$$a^* = \textit{argmax}_{a \in A}U(a)$$

The final form of the utility model can be written as:

\begin{equation}
a^* = \textit{argmax}_{a \in A}\mathbb{E}(\textit{IG}(a)) - \sum_{j}w_j \mathbb{E}(\textit{EH}_j(a)) - \mathbb{E}(\textit{cost}) 
\label{eq:final}
\end{equation}

In Table 1, we reiterate the key properties of practitioners' decision-making when selecting an ethical evaluation practice. We accompany these properties with guiding questions for the ML industry to explore, in order to move towards prioritizing ethics in evaluations. Combining these properties into Equation \ref{eq:final}, we see that a best evaluation practice is chosen after considering the information gained from the practice, potential ethical harms, and is limited by the resources available. 

\begin{table*}[ht]
    \centering
    \begin{tabular}{|p{0.35\textwidth}|p{0.2\textwidth}|p{0.35\textwidth}|}
        \hline
        \textbf{Utility Framing} & \textbf{Explanation} & \textbf{Question for ML community} \\
        \hline
         $\textit{choose } a \in A$ & Selecting an evaluation out of a set of options & \emph{How can we ensure that practitioners are carefully weighing a range of options for evaluation?} \\
        \hline
        $u(a) = \textit{information gain} - 
       \textit{ethical harm} - \textit{cost}$
& Utility is composed of information gained, ethical harms, and costs & \emph{How can we ensure ethical harms introduced in evaluation are considered?} \\
        \hline
         $u(a) = \mathbb{E}(\textit{IG}(a)) - \mathbb{E}(\textit{EH}(a)) - \mathbb{E}(\textit{cost})$ &   Information gained, ethical harm, and costs are in expectation & \emph{How can we ensure practitioners account for potential ethical harms and issues in estimating harms incurred in evaluation?} \\
        \hline
        $ \mathbb{E}(EH(a)) = \sum_{j}w_j \mathbb{E}(\textit{EH}_j(a))$ & Ethical harm is composed of weighted ethical values & \emph{Which specific ethical values are impacted by evaluations? How might regulatory requirements for particular ethical values impact the choice of evaluation?} \\
        \hline
        $u(a^*) = \textit{argmax}_{a \in A_c}U(a)$
& The best evaluation has the highest utility & \emph{How do practitioners who do consider ethical harms define the best evaluation framework and then compare between options?}\\
        \hline
    \end{tabular}
    \caption{Properties for a utility model framing the costs, benefits, and resource constraints for a team's decision-making.}
    \label{tab:tab1}
\end{table*}

The conceptual framework above provides a high-level sketch of how ethical harms associated with machine learning evaluation can affect the overall utility derived from the evaluation process. We now discuss how common challenges practitioners face in the process of selecting an evaluation practice raise questions about whether the best evaluation has been chosen. 

Discussing the interactions between components of the framework, the philosophical challenges, and practical challenges that can arise serves two purposes. First, these challenges do appear in practice. We illustrate the nature of ethical harms resulting from real-world evaluation practices in order to make concrete the sorts of consequences that appear in selecting evaluation practices. We selected the examples below using a broad search across scholarly publications and news media related to ethical issues that arise in ML, with a specific focus on those that can affect evaluation. We focused on identifying instances that varied in evaluation practice, ethical values at risk, and context. We also prioritized examples that provided clear insights into potential incurred harms, or direct evidence of ethical harms. 

Secondly, we discuss real-world evaluations to motivate exploration of mitigation strategies within the ML evaluation industry. We accompany each common challenge with an existing mitigation strategy from ethically-motivated evaluations in domains other than ML. Other fields have established regulatory and administrative systems that help them balance the tradeoffs which arise in evaluation practices, or have informal best practices. These potential mitigation strategies can guide future discussion and move the ML industry towards balancing compliance with ethics.

Our discussion is distilled into the notation from the notation from the utility model in Table 2.

\begin{table*}[ht]
    \centering
    \fontsize{9pt}{9pt}\selectfont
    \begin{tabular}{|p{0.2\textwidth}|p{0.25\textwidth}|p{0.25\textwidth}|p{0.2\textwidth}|}
        \hline
        \textbf{Challenge in balancing evaluation considerations} & \textbf{Example from ML evaluation practices} & \textbf{Mitigation strategies in analogous domain}  & \textbf{Open question for ML community}\\
        \hline
        Ethical harm, $\mathbb{E}(EH(a)$, combines expected ethical harm of many individuals or groups who may be impacted differently & Medical AI device testing done on a single geographic site as a form of `data splitting' can yield low external validity and low utility to diverse populations & The medical community offers best practices to broaden representation in clinical trials and widen external utility through information campaign & How can we ensure $EH(a)$ considers  consider individual and group specific impacts of an evaluation? \\
        \hline
         Subjective interpretations of the presence and magnitude of $\mathbb{E}(EH)$ & Facebook NewsFeed randomized experiment was criticized by some as causing social harm or unfairness during evaluation & IRBs act as a centralized agency and 3rd party overseeing the approval of placebo control trials prior to deployment with a focus on ethics of the evaluation process & How can we ensure decision makers are able to estimate $\mathbb{E}(EH)$ despite conflicting opinions?  \\
        \hline
        
        Utility balances tradeoffs between  future $\mathbb{E}(IG(a))$ from the evaluation and potential ethical harms, $\mathbb{E}(EH(a))$ & Adversarial testing is popular due to its anticipated information gain. But, ethical harms have been established in adversarial testing. For example, guidelines for the data labelling step establish cultural insensitivity or social harm, and other cases of adversarial testing in the physical domain illustrate privacy losses. &  US NEPA guidelines require documentation and consideration of
         environmental harm when proposing federal actions, but is criticized because the documentation is not required to be a primary decision factor. A recommendation to improve environmental impact assessments includes follow-up monitoring.
         & How can the ML industry identify and regulate the consideration of particular ethical harms? \\
        \hline
        Difficulties in comprehensive risk assessments, including unknown probability of ethical harm, $\mathbb{E}(EH(a))$ at time of decision & Microsoft Tay was released for live testing following offline user studies and stress-testing. However, unknown vulnerabilities were not revealed in offline testing, and these led to ethical harm through cultural insensitivity during live testing. & US NRC conducts probabilistic risk assessments on nuclear power plants, focusing on distributions and likelihoods of risks. They impose safety standards to account for uncertainties in the distributions  &  How can we motivate better and more careful assessments of potential ethical harms during the evaluation process? \\ \hline
        $u(a)$ having a negative relationship with $\mathbb{E}(\textit{cost})$ can lead to fewer evaluations than preferred by regulators & Tesla autonomous live testing has been involved in crashes in test driving, causing ethical harm through the value of social harm. Critics recommend raising standards for offline testing, which would require corporations to invest more resources to evaluations & US court decisions in the 1960s held manufacturers responsible for crashworthiness of vehicles, motivating manufacturers to use costly ATDs in testing in offline testing prior to release. & How can we ensure practitioners devote more resources to evaluations? \\
        \hline

        $u(a)$ does not capture the importance of downstream decisions ex-post evaluations & ML applications in education have been criticised for only evaluating model accuracy, rather than the use of models through impact on students in an intervention & Financial regulators impose stress-testing standards, including scenarios with sequences of decisions & How can we ensure that evaluation decisions are downstream actionable, in the face of considerable uncertainty?\\
        \hline

    \end{tabular}
    \caption{
    A summary of challenges in selecting ethical evaluations that are implied by the framework provided. The reality of each of these challenges in the machine learning industry is illustrated by providing an example for each. We also provide an example of a mitigation strategy in an analogous domain to motivate a discussion of lessons for the ML industry. This is not meant to be a comprehensive listing of issues in creating ethical evaluations, but allows us to explore key questions that could motivate avenues to move towards ethical AI through practitioner or regulatory action. }
    \label{tab:tab2}
\end{table*}

\subsection{Issue 1: Aggregating over populations masks group and individual differences.}

Taking the expectation of ``ethical harm" aggregates over individuals and groups. Just as a particular value for a model error metric (like accuracy) or a point estimate (like an estimated average treatment effect) can admit numerous solutions that vary at the level of the individual units or groups (e.g.,~\cite{coston2021characterizing,gelman2023causal,marx2020predictive}), aggregating ethical harms over different individuals can lead evaluators to overlook individual or group-specific concerns. For example, two evaluation protocols may be expected to result in the same level of ethical harm to participants, while differing greatly in how harm is distributed over the specific participants or groups of participants.

\emph{Example in ML Evaluations: Medical AI Device Testing}
Researchers have raised concerns regarding evaluation practices of FDA-approved medical devices. In an analysis of 130 devices, 93 did not have multi-site assessment, meaning many were evaluated at one site, which may have limited geographic diversity. This includes 54 high-risk devices, and devices affecting a range of body areas (chest, breast, heart, head, other) \cite{wu2021medical}.

The evaluation of medical AI devices on limited samples of the population is a form of the general practice of `data splitting', where practitioners partition a population and monitor the performance of the model within a slice of data. This requires careful decision making on the choice of the slice \cite{chen2019slice}. In this case, researchers criticize single-site assessment of medical AI device testing because the relationship between the performance information that is gained and performance in the broader population is unclear. Deploying a method evaluated on a narrow slice of the intended population can yield unintended biases in performance of the device on underrepresented groups post-deployment \cite{wu2021medical}. The FDA has noted these ethical concerns, calling for greater transparency in testing and improved monitoring of algorithmic bias~\cite{wu2021medical}.

\emph{Mitigation Example from Analogous Domain: Representation in Clinical Trials}

The lack of representation in clinical research has been studied in contexts outside of medical devices.  Statistical adjustment techniques, like population-weighted sampling and post-stratification, are common mitigation strategies in the causal inference literature. 

Poor evaluation choices can mask heterogeneity in health needs, leading to downstream harms. For example, not having information on certain subsets of a population may ultimately result in a lack of access to effective interventions for some groups, because sufficient information was not available to obtain treatments. This can potentially compound effects of health disparities, and increase costs \cite{bibbins2022improving}. 

Proposed solutions to widen the inclusivity of clinical trials tend to imply that more resources must be expended in evaluation to improve the value of the information that is gained. These include tailoring recruitment materials with language that emphasizes available support and the value of participant involvement, and providing transportation for participating in a trial \cite{clark2019increasing}.

\emph{Question for ML evaluation industry:} How can we develop evaluation selection guidelines that motivate evaluators to consider individual and group specific impacts of an evaluation design?  

\subsection{Issue 2: Disagreement on presence or magnitude of ethical harms.}

When facing a decision regarding the best possible evaluation practice, estimating and agreeing upon expected ethical harm, $\mathbb{E}(EH)$, is a challenge. Some work in the machine learning and ethics literature argues that a universally acceptable function ranking ethical outcomes does not exist, and that impartiality is simply an ideal \cite{card2020consequentialism}. Practitioners' interpretations of an ethical harm may differ between team members, or with the general public. Furthermore, ethical impacts are considered hard or even impossible to quantify, making it a challenge to prioritize them in metrics-driven development environments \cite{ali2023walking}. This is distinct from the potential issue above, in that the magnitude may be similar across demographic groups but still difficult to agree upon. 

\emph{Example in ML Evaluations: Facebook NewsFeed Randomized Experiment} For one week in January of 2012, Facebook ran a large scale randomized experiment on over 600,000 unknowing participants, randomizing the NewsFeed ranking algorithm they saw. The NewsFeed was the primary mechanism through which individuals saw their friends' content \cite{kramer2014experimental}, and relied on a machine learning algorithm optimized for user behavior while incorporating many weights \cite{meyer2015two}.

This experiment, as with A/B testing more generally, is useful in allowing platforms to observe model performance in complex interactive systems, in order to establish the superiority of one approach over the other with a high degree of statistical certainty. 
The experiment yielded information gain that contributed to internal evaluation of the feed algorithm, as well as 
broader scientific understanding of \emph{emotional contagion}, the impact on an individual's emotion based on exposure to friends' emotions \cite{kramer2014experimental}. While not necessarily intended to motivate direct action, this kind of knowledge gain presumably informs the platform's future design strategy. 
Hence, the information gained carried some utility to Facebook.
This utility diminished, however, by a backlash among the public and some scholars who perceived it as imposing ethical harm on some users who received the negatively manipulated feed. Interestingly, the public backlash conflicted with what was suggested in some prior studies, that users react negatively to others' \textit{positive}, envy-inducing content (e.g.,~\cite{krasnova2013envy}).
Critics decried the practice of randomly placing some individuals in a treatment intervention that they expected (posthoc) to yield less positive emotions, arguing this produced ethical harm through the values of social harm and unfairness. Some critics called for federal agency investigations, and drew parallels to regulations enforced by Internal Review Boards (IRBs) \cite{meyer2015two}. Others however have pointed out that these critiques, like other critiques of platform feed manipulations, exhibit a common “A/B illusion”~\cite{meyer2015two} where a randomized experiment comparing two policies or treatments (A and B) with an unknown rank order in terms of some measure of quality, is deemed less appropriate than simply implementing either A or B for everyone~\cite{meyer2019objecting}.
Hence, the NewsFeed experiment, as an evaluation, led to highly contrasting views on whether inappropriate ethical harm was incurred.

Ultimately, the response to the perceived ethical harms can be thought of as a loss of utility, one that may have been prevented by a different evaluation design. For example, an alternative design might ask users if they want to opt into experimentation. However, this introduces the potential for less information gain, since selection biases come into play, illustrating the difficulty of balancing these concerns when what constitutes a harm can be contested. 

\emph{Mitigation Example from Analogous Domain: Placebo-Controlled Trials and IRB regulations}
Clinical research, including placebo-controlled trials, are regulated by IRBs \cite{polonioli2023ethics, irbfaq}. A/B tests using placebo-controlled trials may be ethically questionable as they deny some participants access to treatment. However, due to the advantages, such as a rigorous test of efficacy, medical practitioners have established guidelines for when placebo trials are appropriate~\cite{millum2013ethics}.  Some of these principles include granting permission when no proven treatment exists for the disorder being studied, and when patients are exposed to at most ``temporary discomfort or delay in relief of symptoms.'' ~\cite{miller2002makes, millum2013ethics}.

The role of the IRB is to evaluate the ethics of a proposed evaluation \cite{polonioli2023ethics}. In our conceptualization, this is analogous to an external evaluation approving that $\mathbb{E}(\mathit{EH})$ meets standards prior to the experiment proceeding. It has been proposed that corporations running online experiments launch internal IRBs or consider external IRBs \cite{polonioli2023ethics}. There have also been calls to increase the transparency of A/B testing or for platforms offering A/B testing to provide ethics training to practitioners to assist their evaluation design \cite{jiang2019s}. Both of these aim to improve ethical impact estimates prior to deploying live experiments.

\emph{Open Question for ML community:} How can evaluators estimate ethical harms in ways that allow for potentially conflicting opinions on the presence or magnitude of harms? 
\subsection{Issue 3: Difficulty of balancing future gains in utility against immediate ethical harms.}

The conceptualization we propose requires balancing expected ethical harm with expected information gain. Even if ethical harm is established (as previously discussed in issues 1 and 2), situations may exist where teams believe it is permissible to ignore potential ethical harms that could occur in evaluation because the information gained through the process could lead to a more socially beneficial downstream ML system.
In the absence of attempts to more carefully weigh concerns against each other, it is easy for model developers to engage in wishful thinking that minimizes more direct and immediate ethical harms incurred in evaluation under the guise of more abstract expected long-term benefits. 

In such cases, a regulatory requirement on ethical values in evaluation outside of ML could potentially mitigate issues. 

\emph{Example in ML Evaluations: Adversarial Testing.}
Adversarial testing in evaluation has become increasingly popular in machine learning. Adversarial testing can be done either prior to the release of models or on released models as part of an iterated deployment process \cite{googadv, shevlane2023model}. In this process (also called `red-teaming'), practitioners intentionally seek out cases where models can behave in ways that would be undesirable. A common manual approach involves individually devising malicious inputs to provide to models, and inspecting the corresponding outputs. This direct intervention approach aims to allow engineers to uncover model failures or vulnerabilities, and identify corrective steps \cite{googadv, shevlane2023model}. 

However, adversarial testing can introduce ethically harmful impact to practitioners, through the value of harmfulness or cultural insensitivity. This can occur during the model output labeling step, which, according to the Google documentation, ``necessarily involves looking at troubling and potentially harmful text or images, similar to manual content moderation" \cite{googadv}. Potential ethical issues in content labeling have been established in other literature \cite{barbosa2019rehumanized}. A risk is that the ethical value of harmfulness is underweighted as a concern when performing adversarial testing, relative to the more nebulous value of the anticipated information gain and subsequent anticipated safety gains of a system.

Another example of potential ethical harm in adversarial testing occurs in computer vision, where researchers test adversarial physical characteristics. Examples include t-shirts that evade computer vision systems that count individuals, or eyewear that evades facial identification systems. These studies have been critiqued for having limited samples when physical testing relative to digital testing. A further critique is the ethical harm of potential privacy losses, such as not blurring faces for the adversarial testers \cite{albert2020ethical}. 

The unclear value of the information gain and potential ethical harms of adversarial testing are in tension with the rise in enthusiasm around adversarial testing. This enthusiasm is even backed by policy; there was recently an executive mandate to establish guidelines and require corporate reporting of performance \cite{whitehouse2023} for companies developing foundation models. Mandates for adversarial testing should be balanced with consideration towards their potential ethical harms. Future work should explore data valuation frameworks for identifying adversarial examples.

\emph{Mitigation Example in Analogous Domain: US government environmental impact standards and values.}
The 1970 National Environmental Policy (NEPA), is a government regulation that was created to ensure ethical values are considered in evaluations. This requires possible environmental effects of actions to be considered and documented by federal agencies. Federal agencies must begin an environmental review process before their final decisions are made, in which they aim to determine if their proposed actions have causal relations to significant environmental effects. NEPA does not mandate the most environmentally sound alternative be chosen in any decision, but requires organizations have knowledge of the impact of decisions. Among other organizations, the Environmental Protection Agency works on overseeing NEPA \cite{citizensnepa}. 

Enforcing documentation and consideration of ethical harms is not universally accepted as a useful practice. Critics say NEPA is weak due to the lack of accurate ex-ante predictions on environmental impact and lack of follow-up monitoring  \cite{karkkainen2002toward}, surfacing concerns about the difficulty of estimating ethical harms. Furthermore, many categorical exclusions exist, initially introduced to acknowledge that not all governmental actions pose environmental risk. Now, critics say this is abused as a loophole to bypass review \cite{foxhill}. A final criticism is that environmental reviews have become time and cost-consuming, and can delay critical clean energy projects, indicating clear dysfunction \cite{meyersohncnn}. This implies that such evaluations have a higher resource cost that is not well balanced by the value of the information gained. Reform efforts balance the need for action and ethical value evaluations, and recommend fewer exclusions, and adaptive mitigation strategies \cite{meyersohncnn, karkkainen2002toward, foxhill}

Overall, the field of sustainability has established mandates that environmental analysis be considered in a decision-making process. This is analogous to mandating that the set of ethical values, $j$, include sustainability.
The specific mechanics of the policy are critical to ensuring ethical development is successfully accomplished, as illustrated by critics of NEPA. 

\emph{Open Question for ML community:} How can ML application industries identify and regulate the consideration of particular ethical harms? 

\subsection{Issue 4: Difficulties in Comprehensive Risk Assessment in Real-World Environments.}

A relevant challenge posed by the consequentialism framework of ethical decision-making processes is that forecasting future ethical well-being and harms across many hypothetical worlds is difficult \cite{card2020consequentialism}. The expectation of $EH(a)$ has to aggregate ethical harm over expected sources of randomness (e.g., stemming from unknown baseline risks of offensive content in content moderation, or unknown, potentially adversarial user behavior after model deployment). This is intensive for practitioners to think about when making decisions, as they may do what they can to prevent harm and vulnerabilities but still experience unanticipated results. 

\emph{Example in ML Evaluations: Microsoft Tay Chatbot Testing.}
In 2016, Microsoft launched a chatbot, Tay, live online. This online live testing and release came after an offline development lifecycle, during which they conducted user studies, and stress-tested the bot under various conditions, to ensure the bot had positive interactions. They hoped releasing Tay online on Twitter would allow them to reach a larger userbase to learn and improve it \cite{leemicrosoft}. However, within 24 hours, it was removed from Twitter, because a vulnerability in the model led to inappropriate words and images from the chatbot \cite{wolf2017we}. 

When the team performed offline testing on Tay, the information they gained led them to believe that it was ready for online interactions through live testing and release with the public on Twitter. At the time of release, the team did not have awareness of the vulnerability that was later exploited to cause harm \cite{leemicrosoft}, through the ethical values of cultural sensitivity, unfairness, and harmfulness. This illustrates the way in which offline user studies were unable to provide sufficient information gain, and in which the anticipated ethical harms were miscalculated. This could be mitigated by better anticipating future risks.

\emph{Mitigation Example in Analogous Domain: Risk Assessment in Nuclear Power Plant Safety}

The US Nuclear Regulatory Commission (NRC) oversees US nuclear power plants, ensuring they ``operate with minimal risk to public health and safety". They use probabilistic risk assessment techniques that explore likelihood of risks, offering a comprehensive approach including potential initiating events, their respective frequencies, and uncertainties. When the distributions of potential outcomes they see have inadequate uncertainties, they impose defenses and safety margins that account for that uncertainty ~\cite{USNRC2022}. 

The NRC notes that there are difficulties in estimating ethical harms. Rather than focusing on the expected values, they focus on estimating the entire distribution of ethically harmful outcomes in their evaluations. This way, they are able to ensure that safeguards exist against a range of potentially harmful outcomes. The approach of the NRC to ensuring that difficulties in estimating outcomes are accounted for is beyond simply improving their risk assessment models; In the face of uncertainty, they impose sufficiently strict safety margins and defenses to account for the uncertainties \cite{USNRC2022}.

\emph{Open Question for ML Community:}  How can we motivate firms to better assess potential ethical harms arising from real-world interactions between AI systems and users or environments, during the evaluation process?

\subsection{Issue 5: Insufficient Resources for Evaluations.}

As discussed above, cost constraints can challenge responsible model development. For example, many generative machine learning models are trained on large, widely available datasets that are believed to be domain-general, then fine-tuned on many small datasets due to the cost of obtaining high-quality, domain-specific data. If sufficient resources aren't devoted to domain-specific testing, the performance observed in an evaluation might appear to be sufficient, but the model might fail dramatically once deployed. Hence, cost constraints can lead to overestimation of the value of information gained. With regulatory or social norms that penalize ethical harms, practitioners can be motivated to devote more resources to investing in ethical evaluation processes to improve systems.

\emph{Example in ML Evaluation: Tesla Autonomous System Live Testing}
 California permits autonomous vehicle manufacturers to evaluate autonomous driving systems on public roads~\cite{dmvpermitholders}. One of the autonomous vehicle systems utilizing this program is the Tesla Autopilot system, which, during training and testing on public roads with other drivers, was involved in several tragic fatal crashes \cite{siddiqui2022}.

The value of information gained in live testing autonomous vehicles has been established in prior literature. In general settings, live testing provides information about performance given real-time rare events. Driving involves many rare events, which could confuse an autonomous navigation system ~\cite{ackerman2017}. Real-world roads contain phenomena that is difficult to simulate in synthetic environments, such as inclement weather, variable road conditions, aggressive or erratic behavior from other cars, and foreign objects~\cite{kalra2016}. Therefore, testing on live roads allows engineers to study instances of systems returning manual control to the human driver~\cite{banerjee2018}. Studying these interventions allows teams to identify failures in perception or control systems~\cite{wang2020safety}. Additionally, teams can receive feedback from passengers of autonomous vehicles, to learn more about the smoothness of the ride and identify issues with service ~\cite{donfro2018}. 

However, ethical harms have occurred with live autonomous vehicle testing, leading critics to note the under-regulation of autonomous vehicle testing, such that corporations might rely on live testing before conducting sufficiently safer offline tests. Some critics advocate for raising standards such that corporations are required to devote more resources to offline testing. For example, proposals include vision tests regarding abilities to recognize surroundings, including cars and pedestrians, prior to live testing \cite{claybrook2018autonomous}, similar to testing norms for allowing people to operate vehicles. Taken together, the need for real world testing, while additionally adhering to the proposed raised standards for offline tests, would substantially increase the resources needed for evaluations.

\emph{Mitigation from Analogous Domain: Crash test reconstruction and liability for ethical harm.}
Regulators and automotive engineers administer tests to evaluate the ``crashworthiness'' of different vehicles. This is motivated by decisions made by American courts in the the late 1960s, when they began to find automotive manufacturers liable for passenger injuries when elements of the car exacerbated the harms experienced by passengers. Regardless of whether accidents were caused by human error, courts argued that the statistically inevitable nature of car accidents meant that manufacturers carried a duty to minimize the consequences of such accidents ~\cite{choi2019crashworthy}.

Crashworthiness is evaluated through crash tests. In a crash test, researchers place a crash dummy in a vehicle, and remotely drive the vehicle into a barrier in order to simulate a crash~\cite{engber2006}. Sensor readings from the vehicle and the crash dummies –along with camera footage of the crash – are then studied to determine the crashworthiness of the vehicle. These tests can be expensive: estimates of the cost of crash dummies range between roughtly \$100,000 and \$1 million USD \cite{automology2021, hall2015, ferris2022}.

Thus, evaluating vehicle crashworthiness requires engineers to navigate a tradeoff between the cost of a crash test and the information it provides. Automotive companies have historically been incentivized to invest resources in crash testing based on standards, such as legal penalties for failure to do so, and consumer preferences for safe vehicles.

\emph{Open Question for ML Community:} How can we motivate practitioners to devote sufficient resources to evaluations despite the need for resource costs to offset the value of the information gain?

\subsection{Issue 6: Impact of evaluations depends on downstream actions.}

Conceiving the value of information gained in an evaluation can be as challenging as forecasting expected ethical harm. This difficulty arises because the information obtained is often instrumental to subsequent decision processes rather than being valuable in isolation.
The value of information gained from an evaluation can be conceptualized in several ways. Evaluators may simply be interested in determining whether the estimated performance falls within an acceptable range. If it does not, the information becomes an input into a subsequent decision problem where the team must consider what actions should be taken to improve the model’s performance or adjust its deployment context.

To expand upon our notation introduced earlier, at the time of designing an evaluation the team can only estimate the information gain given a choice of evaluation $a$, which we denote $\widehat{IG(a)}$. We denote the realized information gain from $a$ as ${IG(a)}$. 
We use $\mathcal{T}$ to denote the set of possible actions that can be taken on the model to improve the ML system after the evaluation is completed.

On completing evaluation $a$, the team chooses the best post-evaluation action that yields the highest utility:

$$t^* = \textit{argmax}_{t \in \mathcal{T}}U(t(IG(a))$$

The \textit{evaluation gain} can be thought of as the gain in utility from taking post-evaluation $t^*$ compared to taking no action, denoted as $t_0$:

$$EG(a)=\mathbb{E}(U(t^*(IG(a))-U(t_0))$$

Ideally, $EG(a)$ could be used in the computation of $a^*$ in Equation \ref{eq:final} in lieu of $IG(a)$, as follows:

\begin{equation}
a^* = \textit{argmax}_{a \in A_c}\mathbb{E}(\textit{EG}(a)) - \sum_{j}w_j \mathbb{E}(\textit{EH}_j(a)) - \mathbb{E}(\textit{cost}) 
\label{eq:issue6}
\end{equation}
but as illustrated here, computing $EG$ contains a number of additional challenges (choosing the utility-maximizing $t^*(\cdot)$, which is dependent on the observed $IG(a)$), and this introduces considerable uncertainty.

\textit{Example in ML Evaluations}: 
ML algorithms experience a rapid growth in education domains, with applications such as predicting student performance and dropout risk \cite{lakkaraju2015machine}, or evaluating postsecondary admissions.
The recent rapid growth of machine learning techniques in education suffers from questions regarding whether these techniques support education principles and goals. Critics highlight ethical concerns with these algorithms, noting negative impacts on historically marginalized students \cite{liu2023reimagining}.

Critics argue that evaluations of these models often prioritize predictive accuracy over their ability to inform effective educational interventions. To better translate predictions into interventions, one recommendation is to frame products as causal inference problems testable through methods like A/B testing \cite{liu2023reimagining}. Incorporating knowledge of potential post-development actions into model creation can lead to more targeted interventions \cite{liu2024actionability}.

\emph{Mitigation Example in Analogous Domain: Stress Testing in Financial Regulation}

To ensure the stability of financial institutions, financial regulators, such as the International Monetary Fund and the US Federal Reserve Bank, enforce various ``stress testing'' frameworks, in order to assess vulnerability and ensure the stability of macroeconomic conditions in the face of plausible, abnormal shocks, such as major changes to exchange rates, or large credit defaults that reduce anticipated cash flows \cite{IMF, FederalReserve2023CCAR}. 

The IMF enforces specific types of evaluations that include scenarios with sequences of decisions. However, regulators note some limitations, including that these requirements impose significant resource costs and expertise by involved parties and suffer computational complexities or data availability \cite{IMF}. In our framing, this illustrates an example of an evaluation that is performed while investigating downstream options, but notes significant costs to doing so.

\emph{Open Question for ML Community:} How can teams ensure that their evaluation decisions are downstream actionable in the face of considerable uncertainty and additional downstream decision-making?

\section{Discussion}

The concept of choosing an evaluation to maximize utility, defined as a sum over expected information gain, ethical harm, and resource costs, encapsulates how we might idealize ethical evaluation. However, the challenges we discuss to this framing illustrate selecting a good evaluation design in practice happens under significant uncertainty, and disagreement, around how to anticipate information gain, ethical harm, costs, beyond what constitutes these quantities in the first place.

According to our economic analogy, there is no politically agreed upon optimal social welfare function for aggregating utility across different individuals' preferences. The existence of subjectivity and ethical value judgements are broadly agreed in the economic literature to be inevitable in scientific analysis. Facing this difficulty, analysts typically proceed in the exercise of examining the consequences of various valuation judgments \cite{samuelson1983foundations}. The decision-makers in a machine learning evaluation practice must also reflect on a range of consequences prior to making final decisions, with the goal of reconciling as much as possible the impacts across a combination of concerns. 
By discussing the consequences of real-life scenarios where value judgements were problematic and mitigations from analogous domains, accompanied by questions for the evaluation industry to use while reflecting on their options, our conceptualization aims to prompt recognition of complex and nuanced values that arise in evaluation decisions. 

The status quo approach to evaluation in research prioritizes sharing the results of evaluations. The pervasive sharing of code, data, and results has been called ``frictionless reproducibility''~\cite{donoho2024data} and used to explain the recent success of machine learning in the world, but a downside is prioritizing the results of evaluations---specifically, the production of point estimates of performance---over richer detail about the evaluation process and how it was selected. Our work highlights how evaluation choices implicate trade-offs between information gains and potential ethical harms under uncertainty, an under-recognized issue in machine learning development.

\subsection{Recommendations from the model} 
Our discussion suggests two broad directions that the software industry could take toward improving decision-making around evaluation trade-offs. 
First, issues 1 through 3 are likely to benefit from developing external review systems, similar to recommendations made for machine learning auditing. Best practices for external audits recommended by \citeauthor{raji2020saving} \shortcite{raji2020saving} and \citeauthor{raji2022outsider} \shortcite{raji2022outsider} include external oversight boards with data access, accreditation for auditors, and registries of ongoing audits. We echo these recommendations and encourage the evaluation industry to similarly move towards exploring effective external review boards. This would be most useful when considering an evaluation that impacts individuals outside of controlled lab experiments, when participants have not consented to participate in the evaluation. In ML evaluation, concerns regarding impact on non-consenting study participants are typically recorded internally or audited externally ex-post the evaluation practice. Instead, we support the community taking a more proactive stance and moving toward designing third-party review boards to plan evaluation practices. 

An independent regulatory body that develops comprehensive risk-assessment frameworks for AI technology could also be beneficial, if these frameworks are able to enforce and capture the potential consequences of AI systems in diverse, unpredictable environments. For instance, the European Union's AI Act introduces a risk-based approach to AI regulation, which could inform the development of ethical risk assessment frameworks for AI evaluation~\cite{veale2021demystifying,novelli2024ai}.

Secondly, we believe that internal decision-making can benefit from further reflection and resources in order to alleviate potential issues 4 through 6. Teams may need to be incentivized to focus more deeply on potential downstream harms (issue 4), adjust their resource allocations toward ethical practices (issue 5), and focus on selecting evaluations that are actionable, linking evaluation outcomes to specific improvement strategies (issue 6).

Resource constraints are relevant because they can impact the ability to refine evaluation techniques to ensure minimal likelihood of ethical harm. Incentives for private companies need to align toward greater investment in ethical internal evaluations. Reviews of internal AI system audits reveal that when recording concerns, internal stakeholders often prioritize regulators' or customers' issues over those of impacted communities, especially if these populations are systematically neglected or underrepresented. This is potentially due to conflicts of interest or efforts to reduce liability risks \cite{raji2022outsider}. It could be analogously possible that evaluations are also prioritizing regulators or customers. Our recommendation is to promote legal or social incentives that encourage corporations to invest in ethical evaluation practices.

Developing a system of governance that dictates approval for evaluations would require working with a wide range of stakeholders beyond practitioners, including legal experts, regulators, and various institutes that currently engage in AI policy. Future work could use case studies to carefully detail evidence of how specific evaluation missteps could have been prevented, and explore options for investigation prior to undergoing the evaluation processes. Practitioners and regulators could be interviewed or surveyed to understand specific weaknesses in their valuation of ethical values. Taken together, further research can allow the ML ethics community to move towards better-planned evaluations.

\subsection{Alternative conceptualizations}
Our conceptualization of ethical evaluation selection is just one possible framing among many.  While we chose it as the most versatile in that it takes as input predicted values of the terms rather than binary information about whether certain thresholds are passed, 
evaluations in practice may sometimes be better described by alternatives.

For example, an alternative conceptualization is to weigh cost explicitly against the other terms. Then, the choice of evaluation is limited to selection within a set of options that are not expected to exceed some maximum allowable cost; i.e.,  choose $a \in A_{c}$ where $A_{c} \subset A$ and for all $a \in A_{c}, cost(a) \leq max(budget)$. Another framing is concerned with ensuring that expected harm is below some threshold, $t_e$, denoted $\mathbb{E}(EH(a))\leq t_e$. This approach, which corresponds to a ``checklist" of potential ethical implications, corresponds to the approach some AI ethicists observe in industry, albeit with mixed feelings on the formalization of ethics in this way \cite{ali2023walking}. 

Our conceptualization emphasizes that evaluation designs are selected under significant uncertainty about the potential value of the information gained and ethical harms and other costs incurred. One issue that arises in practice is a ``cold start'' problem, where prior to running any evaluation, a team may feel unprepared to estimate the relevant terms. 
Addressing the dynamic aspect of evaluation decisions, where some initial evaluation is designed under low information, then subsequent evaluations designed as follow-up conditional on the results, is likely to be important in practice.
When no model evaluation has yet been run, teams may benefit from considering similar models, if available, from other applications or described in the research literature. When choosing subsequent evaluations, teams should weigh the expected information gain against the current knowledge state.
 
\section{Conclusion} We have discussed potential ethical harms due to AI systems that occur due to decisions made in the evaluation process. To separate and categorize various issues in evaluations, we conceptualize the decision problem faced by practitioners when selecting an evaluation. Our conceptualization frames a primary trade-off between the value of information gained in evaluation and the ethical harms and costs of evaluation incurred. We reference best practices for effective evaluations in analogous domains, as well as recommendations made by the machine learning audit research community, to discuss interventions that could improve ethics of evaluations, such as external reviews or devoting additional resources. Our work contributes to the conversation about the need for the machine learning ethics community to focus on deliberately designing evaluations in the development lifecycle to prevent harm from machine learning systems.

\section{Acknowledgements}
We thank the reviewers for their feedback on the paper. We also thank Neel Guha at Stanford University and Yannis Katsis at IBM Research for inputs on the model and examples from analogous domains. This work is supported by the National Science Foundation and the Institute of Education Sciences, U.S. Department of Education through Award $\#2229873$. 

\bibliography{99_refs}

\end{document}